\RequirePackage{lineno}
\documentclass[preprint,tightenlines,showpacs,amsmath,amssymb]{revtex4}
\usepackage{epsfig}
\usepackage{dcolumn}
\usepackage{bm}
\usepackage{color}
\usepackage{graphicx}
\usepackage{subfigure}
\usepackage{pstricks}
\usepackage{pst-node}
\usepackage{rotating}
\usepackage{times}
\usepackage{overpic}

\include{def-com}
\begin{document}
\title{\boldmath Search for the Lepton Flavor Violation Process $J/\psi\to e\mu$ at BESIII}
\author{
M.~Ablikim$^{1}$, M.~N.~Achasov$^{6}$, O.~Albayrak$^{3}$, D.~J.~Ambrose$^{39}$, F.~F.~An$^{1}$, Q.~An$^{40}$, J.~Z.~Bai$^{1}$, R.~Baldini Ferroli$^{17A}$, Y.~Ban$^{26}$, J.~Becker$^{2}$
% ...
\\
\vspace{0.2cm}
(BESIII Collaboration)\\
\vspace{0.2cm} {\it
$^{1}$ Institute of High Energy Physics, Beijing 100049, People's Republic of China\\
$^{2}$ Bochum Ruhr-University, D-44780 Bochum, Germany\\
$^{3}$ Carnegie Mellon University, Pittsburgh, Pennsylvania 15213, USA\\
% ...
$^{6}$ G.I. Budker Institute of Nuclear Physics SB RAS (BINP), Novosibirsk 630090, Russia\\
% ...
$^{17}$ (A)INFN Laboratori Nazionali di Frascati, I-00044, Frascati, Italy; (B)INFN and University of Perugia, I-06100, Perugia, Italy\\
% ...
$^{26}$ Peking University, Beijing 100871, People's Republic of China\\
% ...
$^{39}$ University of Rochester, Rochester, New York 14627, USA\\
$^{40}$ University of Science and Technology of China, Hefei 230026, People's Republic of China\\
% ...
}
}
\vspace{0.4cm}
\begin{abstract}
We search for the lepton-flavor-violating decay of the $J/\psi$ into an
electron and a muon using $(225.3\pm2.8)\times 10^{6}$ $J/\psi$ events collected with the
BESIII detector at the BEPCII collider.
Four candidate events are found in the signal region, consistent with
background expectations.
An upper limit on the branching fraction of
$\mathcal{B}(J/\psi \to e\mu)< 1.5 \times 10^{-7}$ (90\% C.L.) is obtained.
\end{abstract}
\pacs{13.25.Gv, 11.30.Hv, 12.60.-i}
\maketitle
\section{Introduction} \label{sec::introduction}
With finite neutrino masses included, the Standard Model allows
for Lepton Flavor Violation (LFV).
Yet the smallness of these masses leads to a very large suppression,
with predicted branching fractions well beyond current experimental
sensitivity.
However, there are various theoretical models which may enhance
LFV effects up to a detectable level.
Examples of such model predictions,
which often involve super-symmetry (SUSY),
include SUSY-based grand unified theories \cite{ref::susy-gut},
SUSY with a right-handed neutrino \cite{ref::susy-right},
gauge-mediated SUSY breaking \cite{ref::susy-breaking},
SUSY with vector-like leptons \cite{ref::susy-vlepton},
SUSY with R-parity violation \cite{ref::susy-rparity},
models with a Z$^\prime$ \cite{ref::zprime},
and models violating Lorentz invariance \cite{ref::noninvariance}.
The detection of a LFV decay well above Standard Model expectations would be
distinctive evidence for new physics.

Experimentally, the search for LFV effects has been carried out using lepton ($\mu$,$\tau$) decays,
pseudoscalar meson (K,$\pi$) decays,
and vector meson ($\phi$,$J/\psi$,$\Upsilon$) decays, etc.
For example, a recent search for the decay of
$\mu^+\to\gamma e^+$ from the MEG Collaboration
yields an upper limit of
$\mathcal{B}(\mu^+\to\gamma e^+)<2.4\times10^{-12}$ \cite{ref::muon},
and in a similar search with $\tau$ decays the BaBar Collaboration reports
$\mathcal{B}(\tau^+\to\gamma e^+)<3.3\times10^{-8}$ \cite{ref::tau}.
The latest results for neutral kaon and pion decays
from the E871 Collaboration and the E865 Collaboration, respectively,
are
$\mathcal{B}(K^0_L\to\mu^+ e^-)<4.7\times10^{-12}$ \cite{ref::kaon}
and
$\mathcal{B}(\pi^0\to\mu^+ e^-)<3.8\times10^{-10}$ \cite{ref::pion}.
The best $\phi$ decay limit, based on 8.5 pb$^{-1}$ of $e^+e^-$
annihilations at center-of-mass energies from $\sqrt{s}=984-1060$ MeV,
is obtained by the SND Collaboration:
$\mathcal{B}(\phi\to\mu^+ e^-)<2.0\times10^{-6}$ \cite{ref::phi}.
In the bottomonium system, based on about 20.8 million $\Upsilon(1S)$ events,
9.3 million $\Upsilon(2S)$ events, and 5.9 million $\Upsilon(3S)$ events,
the CLEOIII Collaboration presented the most stringent LFV upper limits,
$\mathcal{B}(\Upsilon(1S,2S,3S)\to\mu\tau) < \; \mathcal{O}(10^{-6})$
\cite{ref::cleo_measurement}.
For charmonium, the best limits come from the BESII Collaboration,
who obtained
$\mathcal{B}(J/\psi\to\mu e)<1.1\times10^{-6}$ \cite{ref::bes2-emu},
$\mathcal{B}(J/\psi\to e\tau)<8.3\times10^{-6}$ and
$\mathcal{B}(J/\psi\to\mu\tau)<2.0\times10^{-6}$ \cite{ref::bes2-mutau}
from a sample of 58 million $J/\psi$ events.
A recent sample of 225 million $J/\psi$ events \cite{ref::jpsi_num_inc}
collected with the much improved BESIII detector now allows for LFV searches
in $J/\psi$ decays with a  significant improvement in sensitivity.
We present here our results from a blind analysis of
$J/\psi\rightarrow e^{\pm}\mu^{\mp}$.

\section{BESIII DETECTOR AND MONTE CARLO SIMULATIONS} \label{sec::detector}
The BESIII detector \cite{ref::bes3_detector} at the BEPCII collider
is a large solid-angle magnetic spectrometer with a geometrical acceptance
of 93\% of $4\pi$ solid angle consisting of four main components.
The innermost is a small-cell, helium-based (40\% He, 60\% C$_3$H$_8$)
main drift chamber (MDC) with 43 layers providing an average single-hit
resolution of 135 $\mu$m.
The resulting charged-particle momentum resolution in our 1.0 T magnetic field
is 0.5\% at 1.0 GeV, and the resolution on the ionization energy loss
information ($dE/dx$) is better than 6\%.
Next is a time-of-flight (TOF) system constructed
of 5 cm thick plastic scintillators,
with 176 detectors of 2.4 m length in two layers in the barrel
and 96 fan-shaped detectors in the end-caps.
The barrel (end-cap) time resolution of 80 ps (110 ps)
provides a $2\sigma$ $K/\pi$ separation for momenta up to 1.0 GeV.
Continuing outward, we have an electromagnetic calorimeter (EMC) consisting of
6240 CsI(Tl) crystals in a cylindrical barrel structure and two end-caps.
The energy resolution at 1.0 GeV is 2.5\% (5\%) and the position resolution
is 6 mm (9 mm) in the barrel (end-caps).
Finally, the muon counter (MUC) consisting of 1000 m$^2$
of Resistive Plate Chambers (RPCs) in nine barrel
and eight end-cap layers; it provides a 2 cm position resolution.

Our event selection and sensitivity, including backgrounds,
are optimized through Monte Carlo (MC) simulation.
The {\sc geant4}-based simulation software BOOST \cite{ref::boost}
incorporates the geometry implementation simulations and material composition of the BESIII
detector, the detector response and digitization models
as well as the tracking of the detector running conditions and performances.
The generic simulated events are generated by $e^+e^-$ annihilation into
a $J/\psi$ meson using the generator {\sc kkmc} \cite{ref::kkmc}
at energies around the center-of-mass energy $\sqrt{s}=3.097$ GeV.
The beam energy and its energy spread are set according to measurements
of BEPCII, and initial state radiation (ISR) is implemented
in the $J/\psi$ generation.
The decays of the $J/\psi$ resonance are generated by {\sc evtgen}
\cite{ref::evtgen} for the known modes with branching fractions
according to the world-average values \cite{ref::pdg2012}, and
by {\sc lundcharm} \cite{ref::lundcharm} for the remaining unknown decay modes.

\section{Event selection} \label{sec::analysis}
%
%

We search for events in which $J/\psi$ decays into an electron
and a muon.
Candidate signal events are required to have two well-measured tracks
with $|\cos\theta|<0.8$ and zero net charge,
consistent with originating from the interaction point.
Here, $\theta$ is the polar angle with respect to the beam axis and
the closest approach of each track to the interaction point must
be less then $5$ cm ($1$ cm) in the beam direction
(in the plane perpendicular to the beam).
To reject cosmic rays, the TOF difference between the charged tracks
must be less than 1.0 ns.
The acollinearity and acoplanarity angles between two charged tracks
are required to be less than $0.9^{\circ}$ and $1.4^{\circ}$,
respectively, to reduce other backgrounds.

In order to suppress the radiative events from $e^+e^-\to\gamma e^+e^-$
and $e^+e^-\to\gamma\mu^+\mu^-$, we veto events with one or more good photon
candidates passing the following requirements.
Candidate showers reconstructed in both the EMC barrel region
($|\cos\theta|<0.8$) and in the end-caps ($0.86<|\cos\theta|<0.92$)
must have a minimum energy of 15 MeV.
Showers in the angular range between the barrel and end-cap
are poorly reconstructed and are not considered.
Showers caused by charged particles are eliminated by requiring
candidates to be more than 20 degrees away from the extrapolated
positions of all charged tracks.
Requirements on EMC cluster timing suppress both electronic noise
and energy deposits unrelated to the event.

\begin{figure}[hbt]
\subfigure{ \label{fig7:mini:subfig:a}
\begin{minipage}[b]{0.5\textwidth}
\centering
\includegraphics[width=\textwidth]{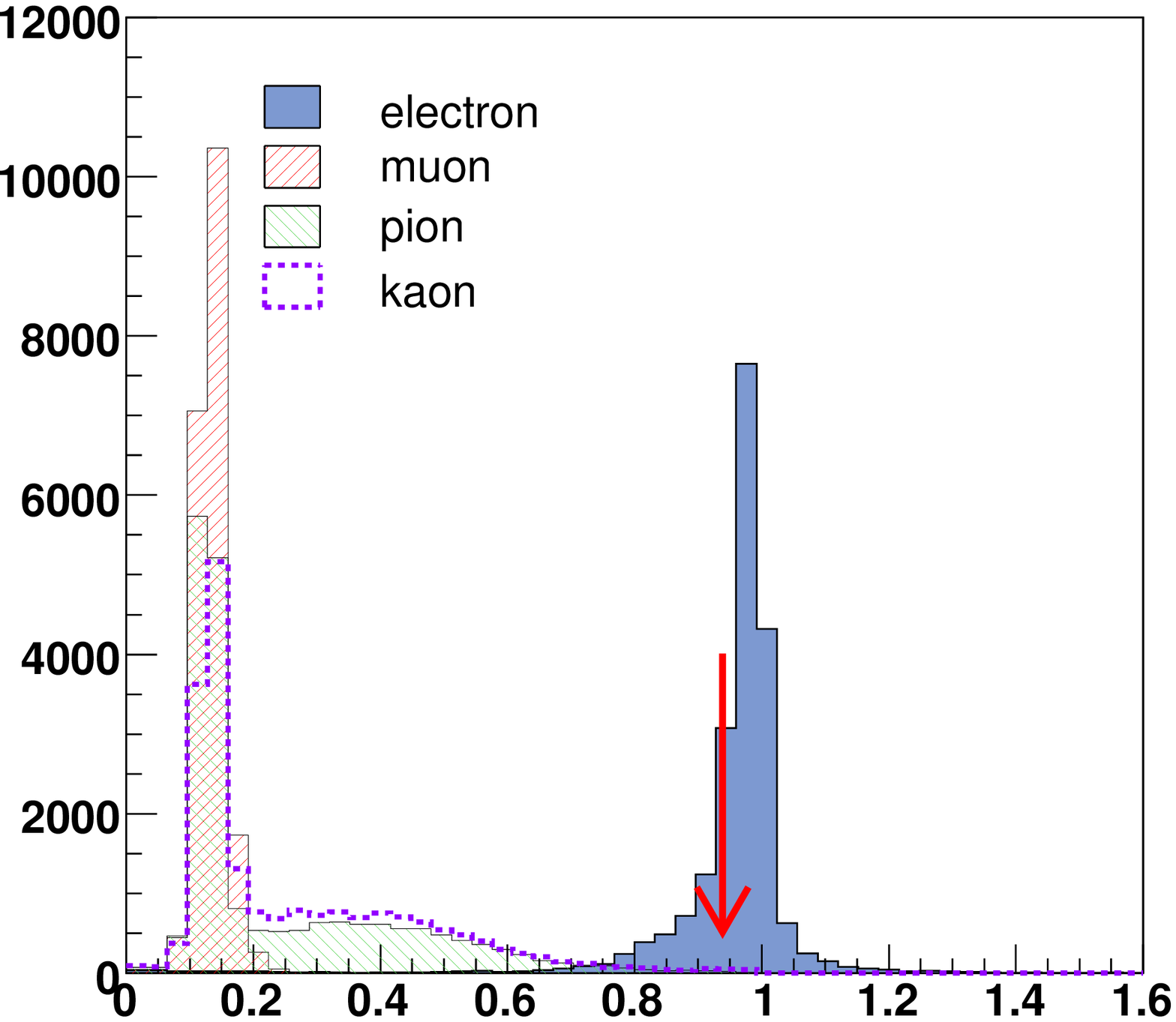}
\put(-145,0){\bf \large $E/p$}
\put(-250,80){\rotatebox{90}{\bf \large number of tracks}}
\put(-85,100){\bf \large (a)}
\end{minipage}}%
\subfigure{ \label{fig7:mini:subfig:b}
\begin{minipage}[b]{0.5\textwidth}
\centering
\includegraphics[width=\textwidth]{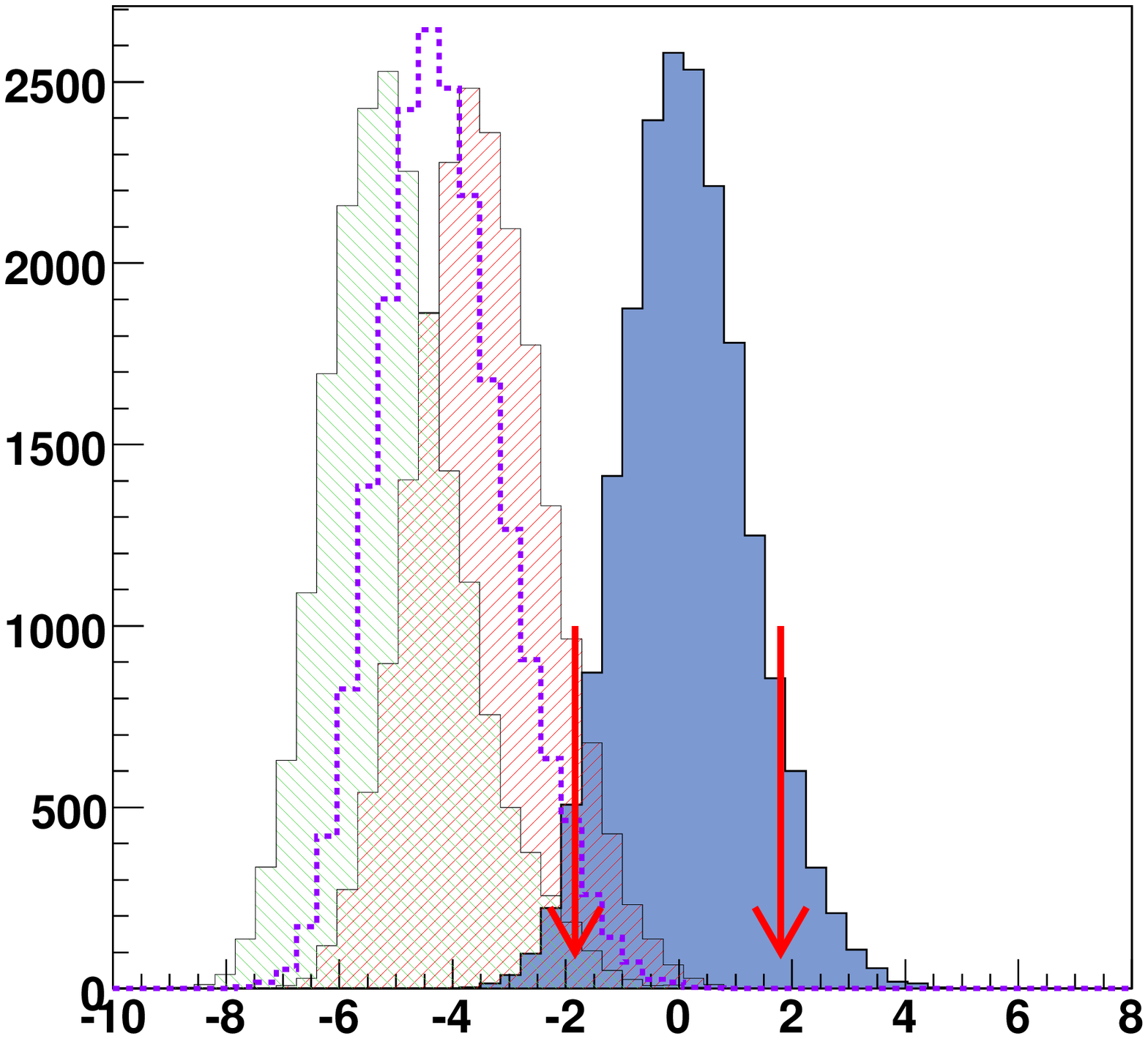}
\put(-145,0){\bf \large $\chi_{dE/dx}^e$}
\put(-250,80){\rotatebox{90}{\bf \large number of tracks}}
\put(-85,100){\bf \large (b)}
\end{minipage}}%
\caption{The distributions of (a) $E/p$ and (b) $\chi_{dE/dx}^e$
for the simulated electron, muon, pion and kaon samples.}
\label{fig7}
\end{figure}

\begin{figure}[hbt]
\subfigure{ \label{fig8:mini:subfig:a}
\begin{minipage}[b]{0.5\textwidth}
\centering
\includegraphics[width=\textwidth]{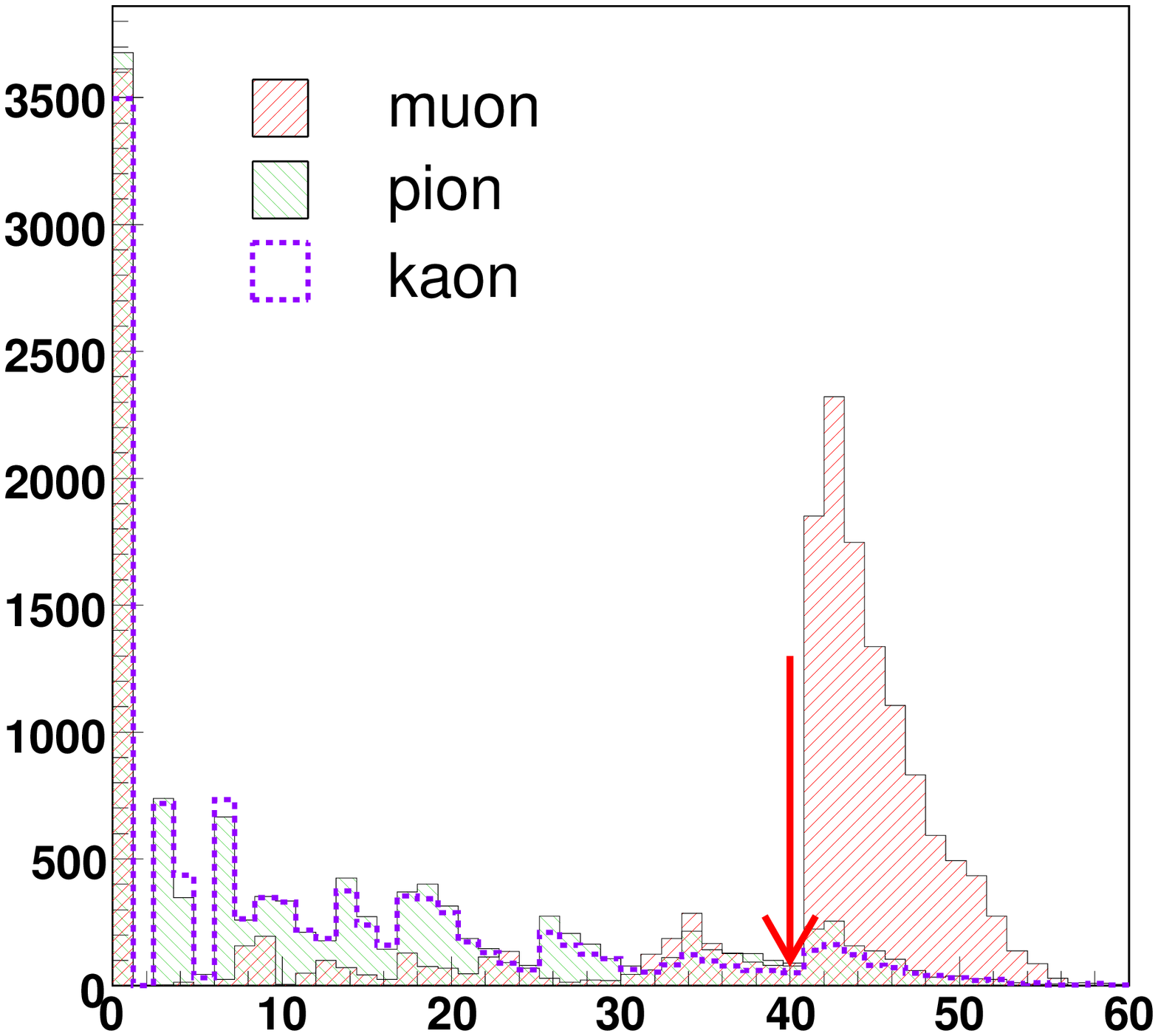}
\put(-145,0){\bf \large Depth (cm)}
\put(-250,80){\rotatebox{90}{\bf \large number of tracks}}
\put(-110,100){\bf \large (a)}
\end{minipage}}%
\subfigure{ \label{fig8:mini:subfig:b}
\begin{minipage}[b]{0.5\textwidth}
\centering
\includegraphics[width=\textwidth]{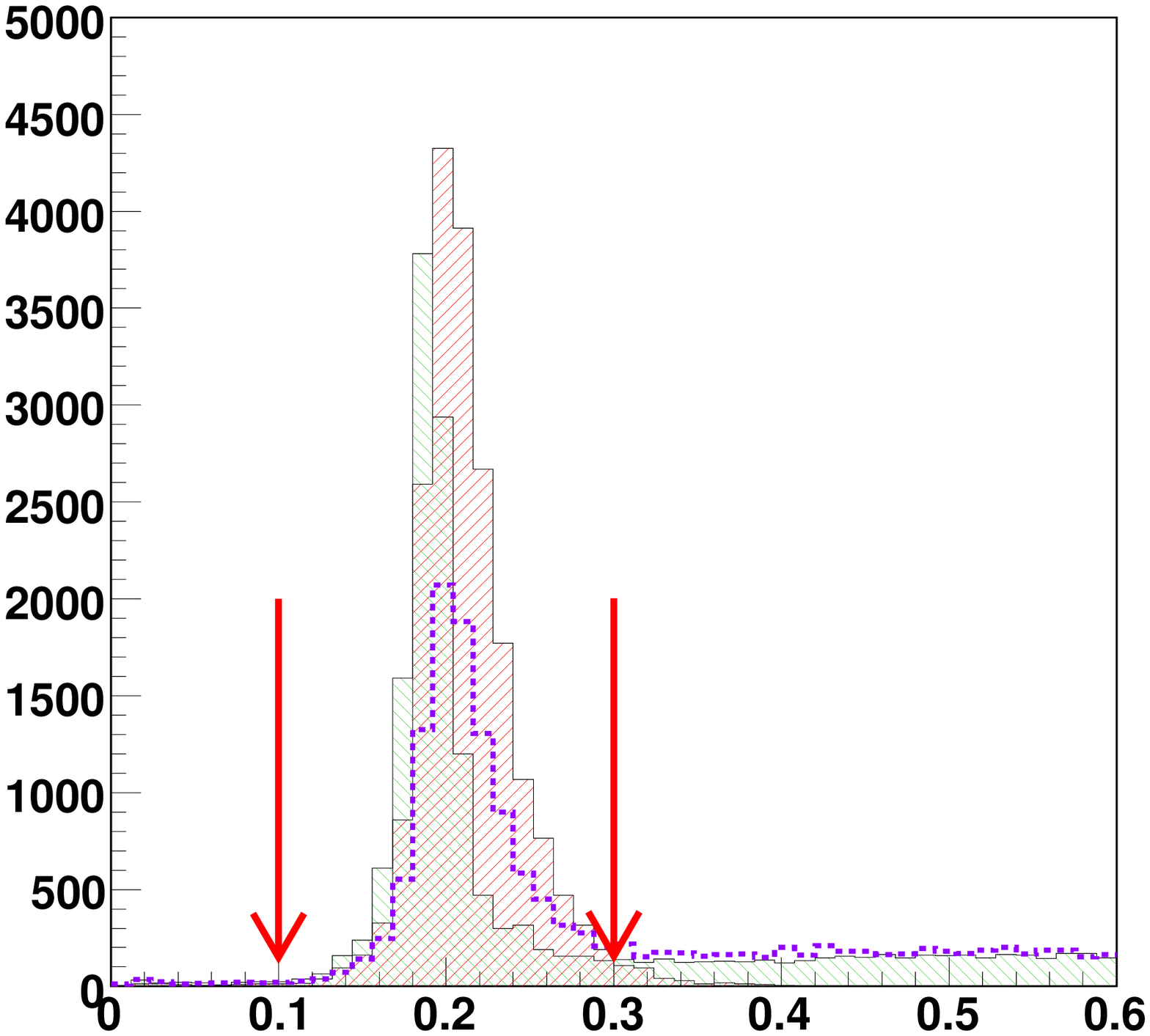}
\put(-145,0){\bf \large Energy (GeV)}
\put(-250,80){\rotatebox{90}{\bf \large number of tracks}}
\put(-110,100){\bf \large (b)}
\end{minipage}}%
\caption{
The distributions of (a) the penetration depth in the MUC and
(b) the deposited energy in the EMC for the simulated muon, pion and kaon
samples.}
\label{fig8}
\end{figure}

The above selection criteria retain events with back-to-back charged tracks
and no obvious extra EMC activity.
Most of the remaining events originate from the background processes
$J/\psi \to e^+e^-, J/\psi \to \mu^+\mu^-,
J/\psi \to\pi^+\pi^-, J/\psi \to K^+K^-,
e^+e^- \to e^+e^- (\gamma)$ and $e^+e^- \to\mu^+\mu^-(\gamma)$,
In order to suppress the these background events, we identify electrons
and muons based on the information of the MDC, EMC and MUC sub-detectors.
The requirements are determined using electron, muon, pion and kaon
samples from $J/\psi \to e^+e^-, \mu^+\mu^-, \pi^+\pi^-, K^+K^-$ MC events.
Electron identification requires no associated hits in the MUC
and $0.95 < E/p < 1.50$, where $E$ is the energy deposited in the EMC
and $p$ is the momentum measured by the MDC.
Also, the absolute value of $\chi_{dE/dx}^{e}$ from comparing the $dE/dx$
measurement with the expected electron signal should be less than 1.8.
Fig. \ref{fig7} shows the $E/p$ and $\chi_{dE/dx}^{e}$ distributions
for electrons, which are well-separated from other particles.
Muon identification uses the barrel MUC system which
covers $|\cos\theta| < 0.75$.
Charged tracks are required to have $E/p < 0.5$ and
a deposited energy in the EMC  $0.1 < E <  0.3$ GeV.
We require the penetration depth in the MUC to be larger than 40 cm;
if the track penetrates more than three detecting layers in the MUC,
we also require the MUC track fit to have $\chi^2 < 100$.
Finally, the $\chi_{dE/dx}^{e}$ value from the $dE/dx$ measurement
calculated with the electron hypothesis must be less than $-1.8$.
The simulated distributions of the deposited energy in the EMC
and the penetration depth in the MUC are shown in Fig. \ref{fig8}.


%
%
Our final analysis of event yields for  $J/\psi \to e^+\mu^-$ is performed
with the two variables $|\Sigma\vec{p}|/\sqrt{s}$ and $E_{vis}/\sqrt{s}$,
where $|\Sigma\vec{p}|$ is the magnitude of the vector sum of the momentum
in the event, $E_{vis}$ is the total reconstructed energy
(calculated using $\sqrt{p^2 + m^2}$ with each track momentum $p$),
and $\sqrt{s}$ is the center-of-mass energy.
Our signal region is defined by $0.93 \leq E_{vis}/\sqrt{s} \leq  1.10$
and $|\Sigma\vec{p}|/\sqrt{s} \leq  0.10$, which correspond in each case
to about two standard deviations as determined by MC simulation.

The analysis is done in a blind fashion in order not to bias our choice
of selection criteria.
Before examining the signal region, all selection criteria were optimized
based on simulated samples with a sensitivity figure-of-merit (FOM)
defined as the average upper limit from an ensemble of experiments
with the expected background and no signal, \
\begin{equation}
\begin{aligned}
  FOM \;=\; \frac{\epsilon}{\sum_{N_{obs} = 0}^{\infty}
  P(N_{obs}|N_{exp})\cdot UL(N_{obs}|N_{exp})} \; ,
\end{aligned}
\end{equation}
where $\epsilon$ is the detection efficiency determined with a sample
of 100,000 simulated $J/\psi\to e\mu$ events,
$N_{exp}$ is the expected number of background events based
on background process simulations,
$N_{obs}$ is the number of observed candidate events,
$P$ is the Poisson probability, and $UL$ is the upper limit on the signal
calculated with the Feldman-Cousins method at 90\% C.L.
\cite{ref::feldman}.
In addition to the signal MC samples, six background MC samples,
each with twice the statistics of the data sample,
are employed to optimize the selection criteria:
$J/\psi \to e^+e^-, J/\psi \to\mu^+\mu^-, J/\psi \to\pi^+\pi^-,
J/\psi \to K^+K^-, e^+e^- \to e^+e^- (\gamma)$,
and $e^+e^- \to\mu^+\mu^-(\gamma)$.

\begin{figure}[htbp]
\begin{center}
\includegraphics[height=8cm,width=10cm]{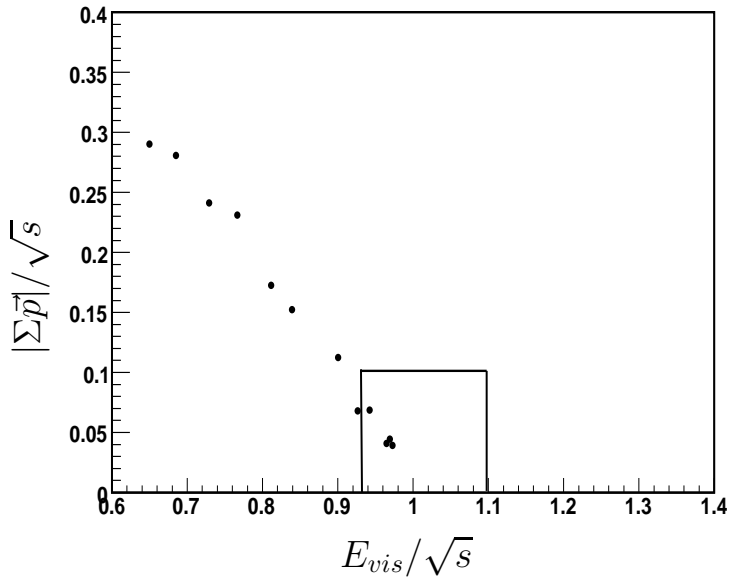}
\put(-170,-5){\bf \large $E_{vis}/\sqrt{s}$}
\put(-295,75){\rotatebox{90}{\bf \large $|\Sigma\vec{p}|/\sqrt{s}$}}
\caption{A scatter plot of $E_{vis}/\sqrt{s}$ versus
$|\Sigma\vec{p}|/\sqrt{s}$ for the $J/\psi$ data.
The indicated signal region is defined as
$0.93 \leq   E_{vis}/\sqrt{s} \leq 1.10$
and $|\Sigma\vec{p}|/\sqrt{s} \leq  0.1$.}
\label{fig::openbox}
\end{center}
\end{figure}

After applying the optimized selections criteria, four candidate events
remain in our signal region, see Fig. \ref{fig::openbox}.
The detection efficiency for signal is determined to be $(18.99\pm0.12)\%$.
Using an inclusive sample of simulated $J/\psi$ decays with four times
the size of our data sample, we find nineteen background events
surviving in the signal region.
This yields a predicted background of $N_{exp}=(4.75 \pm 1.09)$.

\section{Systematic uncertainties} \label{subsec::systematic}
Systematic uncertainties originate from imperfect knowledge of efficiencies
for the electron and muon tracking requirements
electron and muon identification, acollinearity and acoplanarity requirements,
the photon candidate veto,
and the number of $J/\psi$ events.

\subsection{Tracking efficiency} \label{subsubsec::tracking}
Control samples of $\psi^\prime\to \pi^+\pi^- J/\psi,
J/\psi \to e^+e^-, \mu^+\mu^-$ selected from 106 M $\psi^\prime$ data events
and 106 M $\psi^\prime$ inclusive MC events are used to study the possible
differences in the tracking efficiency between data and MC events.
To determine the tracking efficiency of electrons, we select events
with at least three charged tracks.
Two tracks with low momentum, $p<0.5$ GeV/c, and with opposite charge
are interpreted as the pions.
After requiring the recoiling mass opposite these two pions to satisfy
$|M^{recoil}_{\pi^+\pi^-}-3.097| < 10$ MeV,
we obtain $\psi^\prime\to \pi^+\pi^- J/\psi$ candidates.
For the $e^-$ selection, at least one track is required to have a
negative charge, a momentum in the region from 1.0 GeV/c to 2.0 GeV/c,
and a deposited energy in the EMC greater than 1.0 GeV.
With these three tagging tracks, $\pi^{+}$, $\pi^{-}$, and $e^{-}$,
the total number of $e^+$ tracks, $N^0_{e^+}$, can be determined
by fitting the distribution of mass recoiling from the $\pi^{+}\pi^{-}e^{-}$
system, $M_{recoil}^{\pi^{+}\pi^{-}e^{-}}$.
In addition, one can obtain the number of detected $e^+$ tracks, $N^1_{e^+}$,
by fitting $M_{recoil}^{\pi^{+}\pi^{-}e^{-}}$, after requiring
all four charged tracks to be reconstructed.
The tracking efficiency of $e^{+}$ is then obtained
as $\epsilon_{e^+} \;=\; N^1_{e^+} / N^0_{e^+}$.

Similarly, we can obtain the tracking efficiency for $e^{-}$, $\mu^{+}$
and $\mu^{-}$.  The difference between data and MC simulation is found
to be about 1.0\% in each of these four cases,
which is taken as a systematic uncertainty for tracking.

\subsection{Particle identification} \label{subsubsec::epid}
Clean samples of $J/\psi(e^+e^-) \to e^{+}e^{-}$ with backgrounds
less than 1\% selected from data and inclusive MC events are employed
to estimate the uncertainty of the $e^\pm$ identification.
The event selection criteria for this control sample are identical
to those for our signal channel, including two good charged tracks
and no good photon.
The track with higher momentum is required to satisfy the $e^\pm$
identification criteria described previously, and the other track
is used for the $e^\mp$ identification study.

The electron identification efficiency, obtained by comparing
the number of events with and without electron identification
criteria applied on the selected control sample, is defined by:
$\epsilon_{PID} \;=\; N_{evt}(w/~PID) / N_{evt}(w/o~PID)$,
where $N_{evt}$ is the number of events extracted from the control sample.
It is found that the average efficiency difference between data
and MC simulation is 0.62\% for the track momentum range $1.4 - 1.7$ GeV/c,
which is taken as the systematic error for electron identification.
Applying a similar method, we study the systematic error of the $\mu^\pm$
identification using the control sample $J/\psi(e^+e^-) \to \mu^{+} \mu^{-}$.
We apply corrections based on data-MC differences, and a residual
uncertainty of 0.04\% is obtained for the muon identification
in the momentum range $1.4 - 1.7$ GeV/c.

\subsection{ Acollinearity and acoplanarity angles} \label{subsubsec::phi and theta}
A control sample of $J/\psi \to \mu^+\mu^-$ is employed to estimate
the uncertainty due to the acollinearity and acoplanarity angle requirements.
We obtain the corresponding selection efficiency by comparing the number
of events with and without imposing the acollinearity and acoplanarity angle
requirements on the the selected control sample.
We find an efficiency difference between data and MC simulation of 2.83\%,
which is taken as a systematic uncertainty for acollinearity and acoplanarity angle requirements.

\subsection{Photon veto} \label{subsubsec::gamma veto}
We expect no good photon candidates to be present in $J/\psi\to\mu^+\mu^-$,
and therefore choose this channel as a suitable control sample.
The event selection criteria for this control sample are similar to those
in sub-section B. By comparing the numbers of events before and after
imposing the $\gamma$-veto criteria on the selected control sample,
we can obtain the corresponding selection efficiency.
We find that the difference in efficiency between
data and MC simulation is 1.19\%, which is taken as a systematic uncertainty for photon veto.
\vspace{20pt}

The uncertainty in the number of $J/\psi$ is 1.24\% \cite{ref::jpsi_num_inc}.
Table \ref{tab::systematic} summarizes the systematic error contributions
from different sources and the total systematic error is the sum
of individual contributions added in quadrature.
\begin{table*}[ht]
\begin{center}
\caption{Summary of systematic uncertainties (\%).} \label{tab::systematic}
\begin{tabular}{c c}
\hline \hline Sources & Error \\ \hline
$ e ^\pm$ tracking    & 1.00 \\
$\mu^\pm$ tracking    & 1.00 \\
$ e ^\pm$ ID          & 0.62 \\
$\mu^\pm$ ID          & 0.04 \\
Acollinearity, acoplanarity
                      & 2.83 \\
Photon veto           & 1.19 \\
$N_{J/\psi}$          & 1.24 \\
\hline
Total  &  3.65 \\
\hline\hline
\end{tabular}
\end{center}
\end{table*}

\section{Results}
We observe four candidate events with an expected background of
$4.75\pm 1.09$, and therefore set an upper limit on the branching fraction
of $J/\psi \to e\mu$, based on the Feldman-Cousins method with
systematic uncertainties included.
The upper limit on the number of observed signal events at 90\% C.L.,
$N^{UL}_{obs}$, of 6.15 if obtained with  the POLE program \cite{ref::pole}.
Here, the number of expected background events, the number of observed
events,  and the systematic uncertainty are used as the input parameters.
The upper limit on the branching fraction is given by
\begin{equation}
\mathcal{B}(J/\psi\to e \mu) < \frac{N^{UL}_{obs}}{N_{J/\psi}\cdot\epsilon},
\end{equation}
where $N_{J/\psi}$ is the total number of $J/\psi$ events, and $\epsilon$
is the detection efficiency.
Combining, we find a 90\% C.L. upper limit on the branching fraction
of $\mathcal{B}(J/\psi \to e\mu)<1.5 \times 10^{-7}$.

\section{Summary} \label{sec::summary}
Using $225.3\pm2.8\times 10^{6}$ $J/\psi$ events collected with the BESIII detector,
we have performed a blind analysis searching for the lepton flavor violation
process $J/\psi\to e\mu$.
We observe four candidate events, consistent with our background expectation.
The resulting upper limit on the branching fraction,
$\mathcal{B}(J/\psi\to e \mu) < 1.5\times 10^{-7}$ (90\% C.L.),
is the most stringent limit obtained thus far for a LFV effect
in the heavy quarkonium system.

\section{Acknowledgement}
The BESIII collaboration thanks the staff of BEPCII and the computing center for their strong support. This work is supported in part by the Ministry of Science and Technology of China under Contract No. 2009CB825200; National Natural Science Foundation of China (NSFC) under Contracts Nos. 10625524, 10821063, 10825524, 10835001, 10935007, 11125525, 11235011, 11005061; Joint Funds of the National Natural Science Foundation of China under Contracts Nos. 11079008, 11179007, 11079023; the Chinese Academy of Sciences (CAS) Large-Scale Scientific Facility Program; CAS under Contracts Nos. KJCX2-YW-N29, KJCX2-YW-N45; 100 Talents Program of CAS; German Research Foundation DFG under Contract No. Collaborative Research Center CRC-1044; Istituto Nazionale di Fisica Nucleare, Italy; Ministry of Development of Turkey under Contract No. DPT2006K-120470; U. S. Department of Energy under Contracts Nos. DE-FG02-04ER41291, DE-FG02-05ER41374, DE-FG02-94ER40823; U.S. National Science Foundation; University of Groningen (RuG) and the Helmholtzzentrum fuer Schwerionenforschung GmbH (GSI), Darmstadt; WCU Program of National Research Foundation of Korea under Contract No. R32-2008-000-10155-0


\end{document}